\documentclass[english,showpacs,superscriptaddress]{revtex4}
\usepackage[T1]{fontenc}
\usepackage[latin1]{inputenc}
\usepackage{graphicx}

\makeatletter

\usepackage{graphicx}

\usepackage{graphicx}
\usepackage{bm}

\usepackage{babel}
\makeatother
\begin{document}

\title{Local shell-to-shell energy transfer via nonlocal Interactions in
fluid turbulence}

\author{Mahendra K. Verma }

\affiliation{Department of Physics, Indian Institute of Technology, Kanpur 208016,
INDIA}

\author{Arvind Ayyer}

\affiliation{Department of Physics and Astronomy, Rutgers University, Piscataway,
NJ, USA}

\author{Olivier Debliquy }

\affiliation{Université Libre de Bruxelles, Statistical and Plasma Physics, CP231,
Campus Plaine, 1050 Brussels, Belgium}

\author{Shishir Kumar }

\affiliation{Centre for High Energy Physics, Indian Institute of Science, Bangalore
560012, India }

\author{Amar V. Chandra}

\affiliation{Department of Physics, Indian Institute of Technology, Kanpur 208016,
INDIA}

\date{1 March 2005}

\begin{abstract}
In this paper we analytically compute the strength of nonlinear interactions
in a triad, and the energy exchanges between wavenumber shells in
incompressible fluid turbulence. The computation has been done using
first-order perturbative field theory. In three dimension, magnitude
of triad interactions is large for nonlocal triads, and small for
local triads. However, the shell-to-shell energy transfer rate is
found to be local and forward. This result is due to the fact that
the nonlocal triads occupy much less Fourier space volume than the
local ones. The analytical results on three-dimensional shell-to-shell
energy transfer match with their numerical counterparts. In two-dimensional
turbulence, the energy transfer rates to the near-by shells are forward,
but to the distant shells are backward; the cumulative effect is an
inverse cascade of energy. 

\vspace{1cm} 

\end{abstract}


\maketitle

\section{Introduction}

Many equations in physics, e.g., Schrodinger equation and diffusion
equation, are local in real space. Here, to time-advance a variable
at a point, we need the values of the variables and their finite-order
derivatives at the same point. It is well known that incompressible
Navier-Stokes (NS) equation,\begin{eqnarray}
\frac{\partial\mathbf{u}}{\partial t}+(\mathbf{u}\cdot\nabla)\mathbf{u} & = & -\nabla p+\nu\nabla^{2}\mathbf{u},\label{eq:NS}\\
\nabla\cdot\mathbf{u} & = & 0,\end{eqnarray}
is \emph{nonlocal in real space} \cite{Fris:book,LandFlui:book}.
Here $\mathbf{u}$ and $p$ are the velocity and pressure fields respectively,
and $\nu$ is the kinematic viscosity. The nonlocality is due to the
pressure term of Eq. (\ref{eq:NS}), which is obtained by taking the
divergence of incompressible Navier-Stokes (NS) equation \cite{LandFlui:book}\begin{equation}
\nabla^{2}p=-\nabla\cdot\left\{ \mathbf{u}\cdot\nabla\mathbf{u}\right\} .\end{equation}
Hence, \begin{equation}
p(\mathbf{x},t)=-\int\frac{\nabla'\cdot\left\{ \mathbf{u}(\mathbf{x}',t)\cdot\nabla'\mathbf{u}(\mathbf{x'},t)\right\} }{\left|\mathbf{x}-\mathbf{x}'\right|},\end{equation}
which is nonlocal because $p(\mathbf{x},t)$ depends on the velocity
field at $\mathbf{x}'\ne\mathbf{x}$. 

In Fourier space, incompressible NS equation is\begin{eqnarray}
\frac{\partial u_{i}(\mathbf{k})}{\partial t}+\nu k^{2}u_{i}(\mathbf{k}) & = & -\frac{i}{2}P_{ijm}(\mathbf{k})\int\frac{d\mathbf{p}}{(2\pi)^{d}}u_{j}(\mathbf{p})u_{m}(\mathbf{q}),\\
k_{i}u_{i}(\mathbf{k}) & = & 0,\end{eqnarray}
where\[
P_{ijm}(\mathbf{k})=k_{j}\left(\delta_{im}-\frac{k_{i}k_{m}}{k^{2}}\right)+k_{m}\left(\delta_{ij}-\frac{k_{i}k_{j}}{k^{2}}\right),\]
 $\mathbf{k=p+q}$, and $d$ is the space dimensionality. Note that
the factor $-k_{i}k_{j}k_{m}/k^{2}$ of $P_{ijm}(\mathbf{k})$ is
due to the pressure term. To determine $\mathbf{u}(\mathbf{k},t+dt)$
we need the values of field $u_{i}(\mathbf{p})$ where $\mathbf{k-p}$
could be quite large. Hence, incompressible NS equation in \emph{nonlocal
in Fourier space also} \cite{Orsz:Rev,LandFlui:book}. The basic unit
of nonlinear interactions in turbulence, called triad interactions,
involve three vectors $(\mathbf{k},\mathbf{p},\mathbf{q})$ with $\mathbf{k=p+q}$.
These triad interactions yield energy transfers among the participating
modes. The strength of triad interaction is measured using energy
exchanges in the triad.

The energy transfers between two wavenumber shells can be computed
using the triad interactions. One of the key ingredient of Kolmogorov's
phenomenology of turbulence is {}``local'' shell-to-shell energy
transfer. That is, maximum energy is transferred from a wavenumber
shell to the next wavenumber shell. This observation has been verified
in numerical simulations \cite{Zhou:Local,Doma:Local2}. This result
is surprising in view of nonlocal interactions in both real and Fourier
space. In this paper, we compute the strength of triad interaction,
and shell-to-shell energy transfer in incompressible fluid turbulence
using field-theoretic technique.

There have been many attempts in the past to compute the strength
of triad interactions and the energy transfers in fluid turbulence.
Kraichnan \cite{Krai:71} computed these quantities in both 2D and
3D turbulence using {}``almost Markovian Galilean invariant'' turbulence
model. He showed the in 3D 35\% of the total energy transfer across
a unit wavenumber sphere involves triads in which the smallest wavenumber
is more than one-half of the middle wavenumber. Hence, shell-to-shell
energy transfer in 3D turbulence is local in wavenumber space. Later
Domaradzki and Rogallo \cite{Doma:Local2} numerically computed the
above quantities and observed that energy transfers in shells are
always local, but the triad interactions are nonlocal, i.e., triads
having three wavenumbers of very different magnitudes have large magnitudes.
This is succinctly described by Domaradzki and Rogallo as {}``nonlocal
interactions and local energy transfer''. Domaradzki and Rogallo
found their numerical results to be in an excellent agreement with
their own Eddy-damped quasi-normal Markovian (EDQNM) calculation.
They conjectured that the observed energy transfer as being caused
by triads with at least one wavenumber in the energy-containing range.
Ohkitani and Kida \cite{Ohki} analyzed the triad interactions carefully
and concluded that the nonlocal interaction is strong, but the energy
exchange occurs predominantly between comparable scales. They claimed
that the third mode of much larger scale is indifferent to the energy
transfer as if it were a catalizer in a chemical reaction. Zhou \cite{Zhou:Local}
numerically computed the energy transfers using different wavenumber
summation scheme, and found the energy transfers to be local. Waleffe
\cite{Wale} did a similar analysis using a decomposition of the velocity
field in terms of helical modes. Kishida et al. \cite{Kish} used
wavelet basis to address the same problem and obtained similar results.
For review on this topic, refer to Zhou and Speziale \cite{Zhou:RevLocal}.
In the present paper we re-look at some of the above conjectures. 

In all the above papers, turbulent interactions are measured using
a function $S(k|p,q)$ (usually called transfer function) that denotes
the sum of energy transfers from mode $\mathbf{p}$ and \textbf{$\mathbf{q}$}
to mode \textbf{$\mathbf{k}$} \cite{Lesl:book}. Dar et al. \cite{Dar:flux,MKV:PR}
point out that the energy transfer from one shell to another shell
can not be accurately computed using $S(k|p,q)$, essentially because
the third mode of the interaction could lie outside both the shells
under consideration. To overcome this difficulty, Dar et al. \cite{Dar:flux}
modified the above formulation. They used a new function $S(k|p|q)$,
called \emph{mode-to-mode energy transfer rate}, for the energy transfer
from mode \textbf{p} to mode \textbf{k}, with mode \textbf{q} acting
as a mediator, and showed that the shell-to-shell energy transfer
can be correctly computed by this formalism. The computation of $S(k|p|q)$
is done either numerically or using field-theoretic methods. Kishida
et al. \cite{Kish} used similar formalism as Dar et al. \cite{Dar:flux}
for wavelets and numerically computed the shell-to-shell energy transfer
rates. In the present paper we quantify the triad interactions using
$S(k|p|q)$, and compute them using field-theoretic method. We also
calculate the energy transfer rates between wavenumber shells using
a first-order perturbation theory. Our analytic arguments justify
Domaradzki and Rogallo's \cite{Doma:Local2}, Zhou's \cite{Zhou:Local},
and Ohkitani and Kida's \cite{Ohki} numerical results that the turbulent
interaction is nonlocal, but the shell-to-shell energy transfer is
local. 

In this paper we also compute mode-to-mode energy transfer rate $S(k|p|q)$
for space dimension other than 2. For $d=2$, $S(k|p|q)<0$ for most
of $p<k$, unlike 3D case. This property of $S(k|p|q)$ is the reason
for the inverse energy cascade. We find that the transition from backward
to forward energy transfer takes place at $d_{c}=2.25$. 

The organization of our paper is as follows: in Sec. 2 we compute
$S(k|p|q)$ in the inertial range using first-order field theory.
It is shown that nonlinear interactions in incompressible NS are nonlocal.
The nature of $S(k|p|q)$ for 2D and 3D are contrasted. Sec. 3 contains
estimates of the shell-to-shell energy transfer for neighbouring and
distant shells; since the maximal energy transfer takes place between
neighbouring shells, the shell-to-shell energy transfer is said to
be local. In Secs. 4 and 5 we compute shell-to-shell energy transfer
rates in 3D and 2D respectively. Section 4 also contains a comparison
of analytical results with their numerical counterparts. Section 6
contains an elementary discussion on the energy transfer rates in
Burgers turbulence. Section 7 contains conclusions.

\section{nonlocal interactions in incompressible fluid turbulence}

Kraichnan \cite{Krai:71} has computed magnitudes of triad interactions
using transfer function $S(k'|p,q)$ $(\mathbf{k'}+\mathbf{p}+\mathbf{q}=0)$.
In the following discussion we will compute the strength of triad
interaction using Dar et al.'s mode-to-mode energy transfer rate $S(k'|p|q)$
\cite{Dar:flux}\begin{equation}
S(\mathbf{k'}|\mathbf{p}|\mathbf{q})=-\Im\left(\left[\mathbf{k'}\cdot\mathbf{u}(\mathbf{q)}\right]\left[\mathbf{u}(\mathbf{k')}\cdot\mathbf{u}(\mathbf{p})\right]\right)\label{eq:S}\end{equation}
 that represents energy transfer mode from \textbf{$\mathbf{p}$}
to mode $\mathbf{k}$, with mode \textbf{$\mathbf{q}$} acting as
a mediator. Here $\Im$ represents the imaginary part of the argument.
Here we compute the ensemble average of $S$, $\left\langle S(\mathbf{k}'|\mathbf{p}|\mathbf{q})\right\rangle $,
using the standard field-theoretic technique \cite{Krai:65,Lesl:book,McCo:book}.
We expand the $\left\langle S(\mathbf{k}'|\mathbf{p}|\mathbf{q})\right\rangle $
{[}Eq. (\ref{eq:S}){]} to first order in perturbation (see Verma
\cite{MKV:MHD_PRE,MKV:MHD_Flux,MKV:PR} for details). We assume the
flow to be homogeneous and isotropic.

Following the standard field-theoretic procedure, we perform average
of $S(\mathbf{k}'|\mathbf{p}|\mathbf{q})$ and obtain an expression
for the energy transfer rate from mode \textbf{$\mathbf{p}$} to mode
$\mathbf{k}$, with mode \textbf{$\mathbf{q}$} as a mediator\begin{equation}
\left\langle S(k'|p|q))\right\rangle =\frac{T_{1}(k,p,q)C(p)C(q)+T_{2}(k,p,q)C(k)C(q)+T_{3}(k,p,q)C(k)C(p)}{\nu(k)k^{2}+\nu(p)p^{2}+\nu(q)q^{2}},\label{eq:Savg}\end{equation}
 where $C(k)$ is the equal-time correlation function, and $\nu(k)$
is the effective viscosity. The functions $T_{i}(k,p,q)$ are given
by\begin{eqnarray}
T_{1}(k,p,q) & = & kp\left((d-3)z+(d-2)xy+2z^{3}+2xyz^{2}+x^{2}z\right),\label{eq:T1}\\
T_{2}(k,p,q) & = & -kp\left((d-3)z+(d-2)xy+2z^{3}+2xyz^{2}+y^{2}z\right),\\
T_{3}(k,p,q) & = & -kq\left(xz-2xy^{2}z-yz^{2}\right),\label{eq:T9}\end{eqnarray}
 where $d$ is the space dimensionality, and $x,y,z$ are the cosines
of angles between $(\mathbf{p,q}),$ $(\mathbf{k,q})$, and $(\mathbf{k,p})$
respectively.

We take Kolmogorov's spectrum for the correlation function, i. e.,
\begin{equation}
C(k)=\frac{2\left(2\pi\right)^{d}}{S_{d}(d-1)}\frac{K_{Ko}\left|\Pi\right|^{2/3}k^{-5/3}}{k^{d-1}},\end{equation}
 and renormalized viscosity of McComb and Watt \cite{McCoWatt} for
$\nu(k)$ to be\begin{equation}
\nu(k)=\sqrt{K_{Ko}}\nu^{*}\left|\Pi\right|^{1/3}k^{-4/3},\end{equation}
 where $K_{Ko}$ is Kolmogorov's constant, $\Pi$ is the energy flux,
and $\nu^{*}$ is a constant related to the renormalized viscosity
\cite{McCoWatt,MKV:MHD_RG}. Note that in 2D fluid turbulence, $\Pi$
is negative for the wavenumber region with 5/3 spectral index. McComb
and Watt \cite{McCoWatt,MKV:MHD_RG} have computed $\nu^{*}$ using
renormalization technique. Here we take $K_{Ko}=1.6$, $\nu^{*}=0.38$
for 3D, and $K_{Ko}=6.3$, $\nu^{*}=-0.6$ for 2D.

\begin{figure}
\includegraphics[%
  scale=0.5]{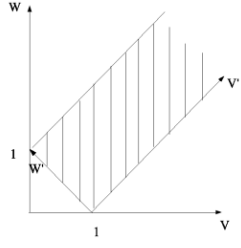}

\caption{\label{Fig:vw} The interacting triad $(\mathbf{k,p,q})/k=(1,v,w)$
under the condition $\mathbf{k}=\mathbf{p+q}$ is represented by a
point $(v,w)$ in the hatched region. The axis $(v',w')$ are inclined
to axis $(v,w)$ by 45 degrees. Note that the local wavenumbers are
$v\approx1,w\approx1$ or $v'\approx w'\approx1/\sqrt{{2}}$. }
\end{figure}

The interactions are self-similar in the inertial range, which is
the region of our interest. Therefore, it is sufficient to analyze
$S(k'|p|q)$ for triangles $(1,p/k,q/k)=(1,v,w)$. Since, $|k-p|\le q\le k+p$,
$|1-v|\le w\le1+v$, hence any interacting triad $(1,v,w)$ is represented
by a point $(v,w)$ in the hatched region of Fig. \ref{Fig:vw} \cite{Lesl:book}.

The local wavenumbers are $v\approx1,w\approx1$, while the rest are
called nonlocal wavenumbers. We substitute $C(k)$ and $\nu(k)$ in
Eq. (\ref{eq:Savg}), which yields \begin{eqnarray}
\left\langle S(v,w)\right\rangle  & = & \left[\frac{4(2\pi)^{2d}K_{Ko}^{3/2}\Pi}{S_{d}^{2}(d-1)^{2}k^{2d}\nu^{*}}\right]*\nonumber \\
 &  & \frac{t_{1}(v,w)(vw)^{-5/3-(d-1)}+t_{2}(v,w)w^{-5/3-(d-1)}+t_{3}(v,w)v^{-5/3-(d-1)}}{1+v^{2/3}+w^{2/3}}\label{eq:Savg_vw}\end{eqnarray}
 where $t_{i}(v,w)=T_{i}(k,p,q)/k^{2}$. For convenience, $\left\langle S(v',w')\right\rangle $
are represented in terms of new variables $(v',w')$ measured from
the rotated axis shown in the figure \ref{Fig:vw}. It is easy to
show that $v=1+(v'-w')/\sqrt{{2}},w=(v'+w')/\sqrt{{2}}$.

\begin{figure}
\includegraphics[%
  scale=0.6]{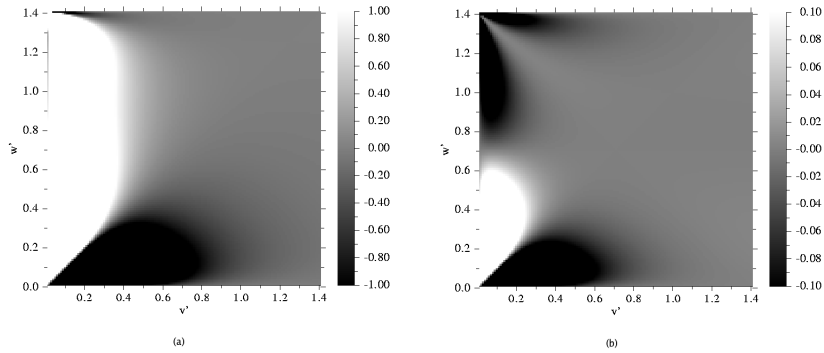}

\caption{\label{Fig:S} Density plot of $\left\langle S(v',w')\right\rangle $
of Eq. (\ref{eq:Savg}) without the bracketed factor for (a) 3D (b)
2D. }
\end{figure}

Fig. \ref{Fig:S} illustrates density plots of $\left\langle S(v',w')\right\rangle $
without the bracketed factor. Fig. (a) shows the plot for 3D, while
Fig. (b) shows the one for 2D. %
Note that $\left\langle S(v'(v,w),w'(v,w))\right\rangle $ is the
energy transferred from mode $p=v$ to mode $k=1$. In the white region
(positive), energy is transferred from mode $p$ to mode $k$, while
in the dark regions (negative), mode $p$ receives energy from mode
$k$. The value of $S$ at $(v,w)=(1,1)$, or $(v',w')=(1/\sqrt{{2}},1/\sqrt{{2}})$
is zero in both 2D and 3D. 

In 3D the triads with $v\approx0,w\rightarrow1$ ($v'\approx0,w'\approx\sqrt{{2}}$:
the top-left corner in $v'$-$w'$ plot) have large and positive $\left\langle S\right\rangle $,
implying that the large wavelength modes give a large amount of energy
to the modes near $k\approx1$. These observations prove that the
nonlinear interactions in incompressible NS \emph{are nonlocal in
Fourier space.} In 2D, the triad with $v\approx0$ have large-negative
$\left\langle S\right\rangle $ implying that the large wavelength
modes take energy from the modes near $k\approx1$. These observations
indicate that the interactions in 2D turbulence are nonlocal as well,
but the large wavelength modes are the sink of energy.

Another common behaviour in both the dimensions is for $w\rightarrow0,v\approx1$
($v'\approx0,w'\approx0$, bottom-left corner in $v'$-$w'$ plot).
Here $\left\langle S\right\rangle \gg0$ for $v<1$, but $\left\langle S\right\rangle \ll0$
for $v>1$. This implies that for these types of triads, the $p$
modes with magnitudes less than $k$ always give energy to the $k$
modes, while the $p$ modes with $p>k$ always take energy from the
$k$ modes. When $p,q$ are much larger than $k$ $\left(v,w\rightarrow\infty\right)$,
$\left\langle S\right\rangle $ is small, implying that they interact
weakly with $k\approx1$. In 3D, $\left\langle S\right\rangle $ for
most of these modes are negative implying that they receive energy
from $k=1$. In 2D, however, $\left\langle S\right\rangle $ for a
large fraction of these triads are positive; hence they supply energy
to $k=1$.

The energy cascade is backward in 2D. This is due to the above mentioned
backward energy transfer from $k=1$ mode to the smaller wavenumber
modes $(\left\langle S(v\approx0,w\approx1)\right\rangle <0)$, and
backward energy transfer from large $v,w$ modes to $k=1$ mode ($\left\langle S(v,w\gg1)\right\rangle >0$).
It is interesting to contrast this behaviour with 3D case where $\left\langle S(v,w)\right\rangle $
is somewhat opposite to 2D case.

\begin{figure}
\includegraphics[%
  scale=0.7]{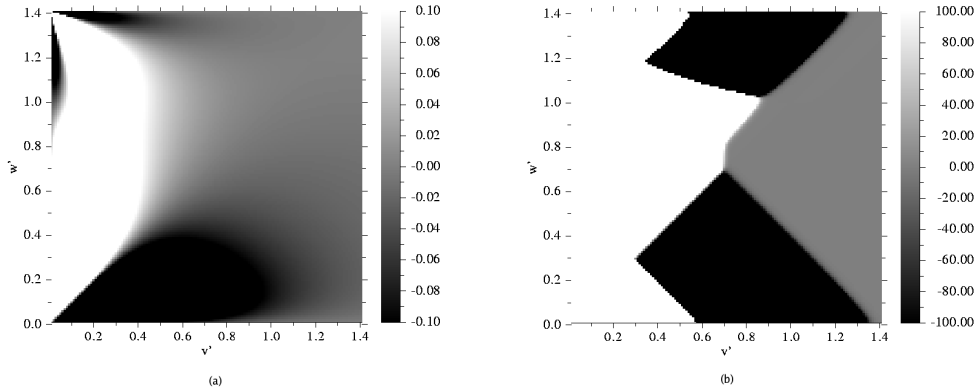}

\caption{\label{Fig:F_d} Density plot of $\left\langle S(v',w')\right\rangle $
of Eq. (\ref{eq:Savg}) for (a) $d=2.25$ and (b) $d=100$. }
\end{figure}

The function $\left\langle S(v,w)\right\rangle $ in the region with
$v\rightarrow0$ is primarily positive for $d=3$, but is negative
for $d=2$. The transition of negative $\left\langle S\right\rangle $
to positive $\left\langle S\right\rangle $ for the region with $v\rightarrow0$
occurs near $d_{c}=2.25$. Please refer to Fig. \ref{Fig:F_d}(a)
for the illustration. It can be shown using field-theoretic calculation
that the renormalized viscosity vanishes near $d_{c}=2.25$, and the
direction of energy cascade changes from negative to positive at $d=d_{c}$.
Fournier and Frisch \cite{FourFris} report $d_{c}=2.05$ which differs
a bit from our $d_{c}$. The difference could be because of the fact
that Fournier and Frisch \cite{FourFris} use combined energy transfer
$S(k|p,q)$ for their EDQNM calculation. %
In Fig. \ref{Fig:F_d}(b) we also show the density plot of $\left\langle S(v,w)\right\rangle $
for $d=100$. This is a representative illustration for large space
dimensions.

The function $\left\langle S(v,w)\right\rangle $ can be estimated
in the limiting cases using the method given in Appendix of Leslie
\cite{Lesl:book}. In the nonlocal region with $v\rightarrow0$ (naturally,
$w\approx1$),\begin{eqnarray}
\left\langle S(v,w)\right\rangle  & \propto & v^{\frac{4}{3}-d}.\end{eqnarray}
 Clearly, $S(v,w)\rightarrow\infty$ for both 2D and 3D. This observation
is consistent with our earlier observation that interactions are nonlocal\emph{.}
However when $v\approx w\approx1$, we find that \begin{equation}
\left\langle S(v,w)\right\rangle \propto\left\{ \begin{array}{c}
-(v-1)\,\,\,\,\textrm{ for }d=3\\
\frac{8}{3}\left[(v-1)^{2}-\frac{1}{2}(w-1)^{2}-\frac{1}{2}(v-1)(w-1)\right]\,\,\,\textrm{for $d=2$ }\end{array}\right.\end{equation}
 This result shows that the interactions within the local triads ($v\approx1,w\approx1$)
are weak. In 3D, among these triads, the ones with $v<1$ or $p<k$
have $\left\langle S\right\rangle >0$, hence the energy transfer
is from mode $p$ to mode $k$; the sign of $\left\langle S\right\rangle $,
consequently the sign of energy transfer, is reversed for the modes
with $v>1$. However in 2D, $S$ is somewhat complex in the neighbourhood
of $v=w=1$.

For $v\approx1$ and $w\rightarrow0$, \begin{eqnarray}
\left\langle S(v,w)\right\rangle  & \propto & w^{\frac{1}{3}-d},\end{eqnarray}
 which again diverges. When $v,w\rightarrow\infty$,

\[
\left\langle S(v,w)\right\rangle \propto v^{-\frac{4}{3}-d}.\]
 implying that interactions with large wavenumber modes are weak. 

The above estimates are consistent with the graphical plots shown
in Figs. \ref{Fig:S} and \ref{Fig:F_d}. After this discussion, we
move on to compute the shell-to-shell energy transfer rates in fluid
turbulence.

\section{Local Shell-to Shell Energy Transfer in Incompressible Fluid Turbulence}

The wavenumber space is divided into shells $(k_{0}s^{n},k_{0}s^{n+1})$,
where $s>1$, and $n$ can take both positive and negative values.
The energy transfer rate from $m$th shell $(k_{0}s^{m},k_{0}s^{m+1})$
to $n$th shell $(k_{0}s^{n},k_{0}s^{n+1})$ is given by \cite{Dar:flux}\begin{equation}
T_{nm}=\sum_{k_{0}s^{n}\leq k\leq k_{0}s^{n+1}}\sum_{k_{0}s^{m}\leq p\leq k_{0}s^{m+1}}\left\langle S(k|p|q)\right\rangle .\label{eq:shell}\end{equation}
 If the shell-to-Shell energy transfer rate is maximum for the nearest
neighbours, and decreases monotonically with the increase of $|n-m|$,
then the shell-to-shell energy transfer is said to be local.

If the amplitudes of the Fourier modes $\mathbf{u}(\mathbf{k})$ are
available, either from experiments and or from numerical simulations,
then we can easily compute the shell-to-shell energy transfer rates
using Eqs. (\ref{eq:S}, \ref{eq:shell}). In this paper, we compute
the energy transfer rates between the wavenumbers shells to first
order in perturbation \cite{Lesl:book,MKV:PR} that yields\begin{eqnarray}
T_{nm} & = & \int_{k_{0}s^{n}\leq k\leq k_{0}s^{n+1}}\frac{d\mathbf{k}}{(2\pi)^{d}}\int_{k_{0}s^{m}\leq p\leq k_{0}s^{m+1}}\frac{d\mathbf{p}}{(2\pi)^{d}}\nonumber \\
 &  & \,\,\,\,\,\,\frac{T_{1}(k,p,q)C(p)C(q)+T_{2}(k,p,q)C(k)C(q)+T_{3}(k,p,q)C(k)C(p)}{\nu(k)k^{2}+\nu(p)p^{2}+\nu(q)q^{2}}\end{eqnarray}
 We nondimensionalize the above equations using \cite{Lesl:book}\begin{equation}
k=\frac{a}{u};\,\,\,\,\,\, p=\frac{a}{u}v;\,\,\,\,\,\, q=\frac{a}{u}w,\end{equation}
 where $a$ is an arbitrary constant wavenumber. For our calculation
we choose $a=k_{0}s^{n-1}$. Three dimensional integral under the
constraint that $\mathbf{k'+p+q}=0$ is given by \cite{Lesl:book}\begin{equation}
\int_{\mathbf{p}+\mathbf{q}+\mathbf{k}=0}d\mathbf{p}=S_{d-1}\int dpdq\left(\frac{pq}{k}\right)^{d-2}\left(\sin\alpha\right)^{d-3}.\label{eq:volume-integral}\end{equation}
 Using these substitutions, we obtain \begin{equation}
\frac{T_{nm}}{\left|\Pi\right|}=K_{u}^{3/2}\frac{4S_{d-1}}{(d-1)^{2}S_{d}\nu^{*}}\int_{s^{-1}}^{1}\frac{du}{u}\int_{us^{m-n}}^{us^{m-n+1}}dv\int_{|1-v|}^{1+v}dw\left(vw\right)^{d-2}\left(\sin{\alpha}\right)^{d-3}F(v,w),\label{eq:shell_final}\end{equation}
 where $F(v,w)$ is given by \begin{equation}
F(v,w)=\frac{t_{1}(v,w)(vw)^{-\frac{2}{3}-d}+t_{2}(v,w)w^{-\frac{2}{3}-d}+t_{3}(v,w)v^{-\frac{2}{3}-d}}{(1+v^{2/3}+w^{2/3})}\label{eq:F}\end{equation}
 with $t_{i}(v,w)=T_{i}(k,kv,kw)/k^{2}$. Eq. (\ref{eq:shell_final})
provides us with the shell-to-shell energy transfer rates with relative
to energy flux $\Pi$. Clearly, $T_{nm}$ depends only on $n-m$,
or $T_{nm}=T_{n-i,m-i}$ where $i$ is an integer. Hence, the shell-to-shell
energy transfer is self-similar.

Now let us estimate the shell-to-shell energy transfer rates when
$m\ll n$. The triads $(1,v,w)$ with $v\rightarrow0$ participate
in this energy transfer. As seen in previous section, for these triads
$S(v,w)\propto w^{(4/3)-d}$. Therefore,\begin{eqnarray*}
T_{nm} & \sim & \int\frac{du}{u}\int dvv^{d-2}v^{\frac{4}{3}-d}\int_{|1-v|}^{1+v}dww^{d-2}\\
 & \sim & v_{0}^{4/3},\end{eqnarray*}
 where $v_{0}$ is the outer radius of shell $m$. Since $v_{0}\rightarrow0$,
$T_{nm}$ vanishes. Hence, we are able to show that shell-to-shell
energy transfer between distant shells in negligible in spite of large
interactions between distant wavenumbers. This result is essentially
due to small volume of $v$ or $m$-th shell.

We can also compute the shell-to-shell energy transfer rates for close-by
shells. Here the participating triads will satisfy $v\approx w\approx1$.
As shown in the earlier section, $S(v,w)\approx-(v-1)$ for these
triads. Therefore, \begin{equation}
T_{nm}\propto\int\frac{du}{u}\int dv\int dw(vw)(v-1),\end{equation}
 with $n\approx m$. Since our bins are uniform in logarithmic scale,
the volume of wavenumber shells is of the order of 1 when $v\approx w\approx1$.
Also the range of $v-1$ is of the order of $v$, which is close to
1. Therefore $T_{nm}$ will be finite. Hence, the shell-to-shell energy
transfer rates between close-by shells is finite. This result is consistent
with the local energy transfer assumption of Kolmogorov.

For close-by shells, $p\approx k$, but $q$ can take any value from
$|k-p|$ to $k+p$. However, it can be easily shown that the mode
$p$ in the triads with $q\ll k$ do not contribute significantly
to the shell-to-shell energy transfer. Since, $S(v,w)\propto w^{\frac{1}{3}-d}$
for $w\rightarrow0$,

\begin{eqnarray}
T_{nm} & \sim & \int\frac{du}{u}\int d\mu\mu\int dww^{d-2}w^{\frac{1}{3}-d}w\\
 & \sim & w_{0}^{1/3},\end{eqnarray}
 where $v=1+\mu w$, and $dv=\mu dw$. Clearly the above integral
goes to zero. That is, the shell-to-shell energy transfer between
close-by shells receive insignificant contribution from the triads
satisfying $q\ll p\approx k$.

In this section we showed using the limiting values of $S(k|p|q)$
that the shell-to-shell energy transfer is local in spite of nonlocal
interactions among wavenumber modes. However, $T_{nm}$ can be easily
computed for the shells in the inertial range. In the next section
we will do these calculations.

\section{Computation of shell-to-shell energy transfer in 3D}

\begin{figure}
\includegraphics[%
  scale=0.8]{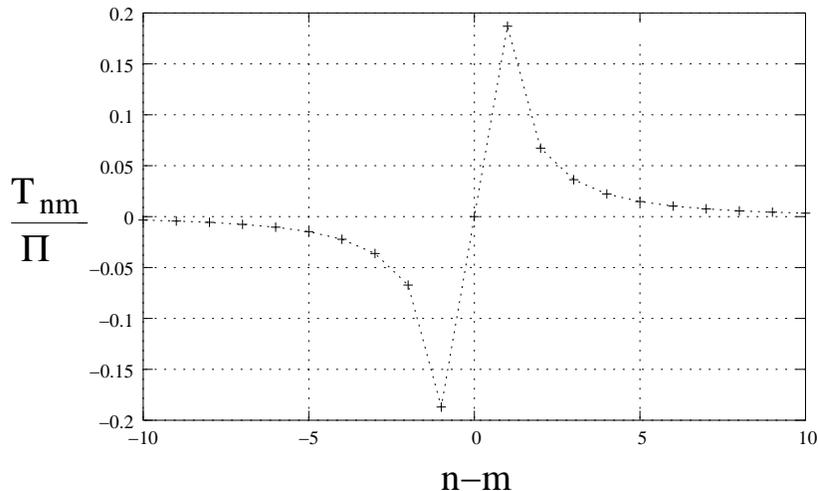}

\caption{\label{Fig:shell3D} Plot of normalized shell-to-shell energy transfer
$T_{nm}/\Pi$ vs $n-m$ for $d=3$. The $n$th shell is $(k_{0}s^{n},k_{0}s^{n+1})$
with $s=2^{1/4}$. The energy transfer is maximum for $n=m\pm1$,
hence the energy transfer is local. The energy transfer is also forward.}
\end{figure}

The shell-to-shell energy transfer rates have been calculated earlier
by Ohkitani and Kida \cite{Ohki}, and Zhou and Speziale \cite{Zhou:RevLocal}
using numerical simulation and EDQNM approximation with $S(k|p,q)$.
In this section we compute the shell-to-shell normalized energy transfer
rates $T_{nm}/\Pi$ in three dimensions using Eq. (\ref{eq:shell_final}).
We take $s=2^{1/4}$. The integration has been done numerically using
Gauss-quadrature method. The constants $\nu^{*}=0.38$ and $K=1.6$
have been taken from McComb and Watt \cite{McCoWatt} and Verma \cite{MKV:MHD_RG,MKV:PR}.
The shell-to-shell energy transfer is self-similar, i. e., $T_{nm}$
is function of $n-m$. Therefore we compute $T_{nm}/\Pi$ for various
$n-m$. Fig. \ref{Fig:shell3D} contains this plot. Note that the
shells $m$ and $n$ have been assumed to be inside the inertial range.

From Fig. \ref{Fig:shell3D} we can infer that the transfer rates
$T_{nm}$ in the inertial range are negative for $n<m$, and positive
for $n>m$. Hence a shell gains energy from the smaller wavenumber
shells, and loses energy to the higher wavenumber shells. This means
that the energy cascades from the smaller wavenumbers to the higher
wavenumbers (forward). The most significant energy transfer takes
place from $m$ to $m+1$. Hence, the shell-to-shell energy transfer
is forward and local, consistent with Kolmogorov's picture of turbulence.
Note that the energy transfer is local in spite of nonlocal triad
interactions.

\begin{figure}
\includegraphics[%
  scale=0.5]{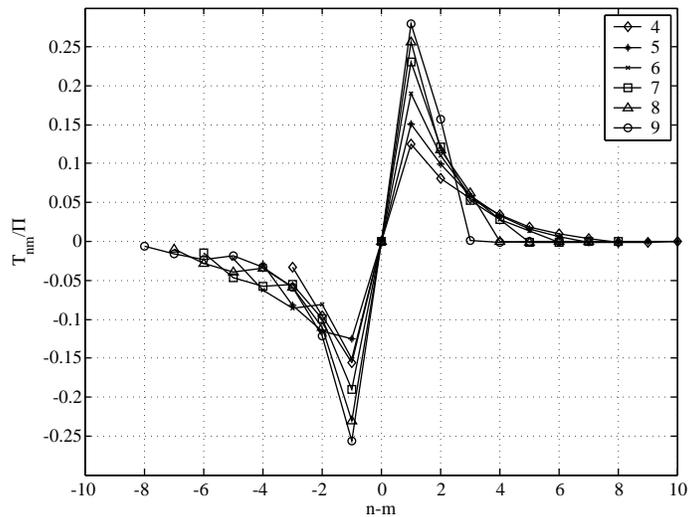}

\caption{\label{Fig:T-sim} Plots of normalized shell-to-shell energy transfer
$T_{nm}/\Pi$ vs. $n-m$ for $m$ from 4 to 9. The plots collapse
on each other indicating self-similarity.}
\end{figure}

To validate our theoretical calculations, we have also computed the
shell-to-shell energy transfer rates using the data from Direct Numerical
Simulation on a $512^{3}$ grid. The computation was performed when
the turbulence was well developed. The Reynold's number using Taylor's
microscale was $64.8$, and skewness was $-0.54$. We divide the wavenumber
space into 15 shells with boundaries at wavenumbers $(2,4,8,11.3,13.5,16,19,22.6,26.9,32,38.1,64,76.1,108,128,256)$.
In the inertial range ($k\approx10-35$) , the shell boundaries are
$k_{n}=2^{(n+11)/4}$. Please refer to Dar et al. \cite{Dar:flux}
for details on numerical procedure. Fig. \ref{Fig:T-sim} shows the
plots of $T_{nm}/\Pi$ vs. $n-m$ for $m=4..9$, which are in the
inertial range shells. %
The plots show self-similarity, local, and forward energy transfer
for the inertial range shells. The  numerical and theoretical values
are in close agreement.

For thicker shells $s=2^{1/2}$, the ratio of the smallest to largest
wavenumber of the triad is $2\sqrt{{2}}$. For these shells, energy
transfer to the nearest neighbouring shell is close to $35\%$. These
numbers are consistent with Kraichnan \cite{Krai:71} and Zhou's \cite{Zhou:Local}
results. 

In the next section we will discuss shell-to-shell energy transfer
in 2D turbulence.

\section{shell-to-shell energy transfer in 2D Fluid Turbulence}

\begin{figure}
\includegraphics[%
  scale=0.6]{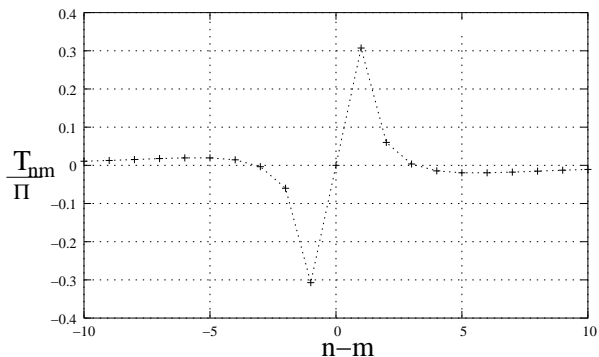}

\caption{\label{Fig:shell2D} Plot of normalized shell-to-shell energy transfer
$T_{nm}/\left|\Pi\right|$ vs $n-m$ for $d=2$ in the inertial range.
The energy transfer rate from the shell $m$ to the shells $m+1,$
$m+2$, $m+3$ is forward, but $m+4$ onward it is negative. The net
effect of all these transfer is the inverse energy flux $\Pi$. }
\end{figure}

We compute the shell-to-shell energy transfer in 2D following the
same procedure as given above. The wavenumber range considered is
in the inverse cascade regime ($E(k)\propto k^{-5/3}$). We take $\nu_{*}=-0.6$
and $K=6.3$ \cite{Krai:71,MKV:PR}. As shown in Fig. \ref{Fig:shell2D},
the energy transfer rates from the shell $m$ to the three neighbouring
shells ($m+1,m+2,m+3$) are \emph{forward}, and the transfers are
negative for all shells $n>m+3$. The above result is very similar
to Dar et al. \cite{Dar:flux}'s numerical finding on 2D MHD turbulence
(Fig. 10 of Dar et al. \cite{Dar:flux}). The negative energy transfer
from the distant shells are due to negative $\left\langle S(v,w)\right\rangle $
for $v\rightarrow0$ (top-left of Fig. \ref{Fig:S}(b)), and positive
$\left\langle S(v,w)\right\rangle $ for $v,w\gg1$. The negative
$\left\langle S(v,w)\right\rangle $ for $v\rightarrow0$ indicate
that large-wavelength modes receive energy from mode $k=1$, and positive
$\left\langle S(v,w)\right\rangle $ for $v,w\gg1$ indicate that
small-wavelength modes give energy to mode $k=1$. A careful inspection
of Fig. \ref{Fig:S}(b) indicates that the forward energy transfer
to shells ($m+1,m+2,m+3$) is due to a narrow region near $v=w=1$,
or $v'=w'=1/\sqrt{{2}}$, where energy transfer is from lower wavenumber
to higher number. 

The above results on shell-to-shell energy transfer is consistent
with the energy flux picture. Note that \[
\Pi=\sum_{n=m+1}^{\infty}(n-m)T_{nm}.\]
 When we perform the $\sum_{n=m+1}^{\infty}(n-m)(T_{nm}/|\Pi|)$,
we obtain $-1$, consistent with the inverse cascade of energy in
2D turbulence. 

To summarize, in 2D fluid turbulence, the shell-to-shell energy transfer
to the neighbouring shells in forward, but the energy transfer is
backward for the distant shells. The above behaviour is due to forward
local transfer and backward nonlocal transfers described in Section
2.

In the next section we will contrast the energy transfers in incompressible
fluid turbulence with relative to Burgers turbulence (compressible
limit).

\section{Locality Issues in Burgers Turbulence}

As discussed in the introduction, Navier-Stokes equation is nonlocal
in real space due to pressure. In Burgers equation\[
\frac{\partial\mathbf{u}}{\partial t}+(\mathbf{u}\cdot\nabla)\mathbf{u}=\nu\nabla^{2}\mathbf{u}\]
the pressure term is dropped with an implicit assumption that the
flow velocity is much greater than sound speed, or the sound speed
is very small. This is the opposite limit of incompressible NS where
the sound speed is infinite. Hence a very different behaviour is expected
for Burgers equation \cite{Fris:book}. Clearly, to time-advance the
velocity field of Burgers equation at a point, we need local field,
and its first and second derivative. Hence, Burgers equation is \emph{local
in real-space}, and nonlocal in Fourier space.

The formula for mode-to-mode energy transfer $S(k'|p|q)$ (Eq. {[}\ref{eq:S}{]})
is not applicable for Burgers equation because it is compressible
$(\nabla\cdot\mathbf{u}\ne0)$ \cite{Dar:flux,MKV:PR}. Therefore,
the shell-to-shell energy transfer cannot be computed accurately.
Note however that energy flux can be computed for Burgers equation.
The energy flux is multifractal, and $\Pi(k)\propto k^{-1/2}$. Therefore,
$E(k)\propto\Pi^{2/3}k^{-5/3}=k^{-2}$ \cite{Fris:book,MKV:KPZ}.

\section{Conclusions}

It is known that the nonlinear interactions in \emph{incompressible}
Navier-Stokes equation is nonlocal in real space due to the pressure
term. In this paper we investigated locality in Fourier space by computing
the strength of triad interactions using the formula for the mode-to-mode
energy transfer. Our calculation is based on first-order field-theoretic
technique. We take Kolmogorov's 5/3 powerlaw for the energy spectrum,
and the renormalized viscosity for the effective viscosity. It has
been shown that the magnitudes of interactions for the nonlocal triads
$k\approx p\gg q$ and $k\approx q\gg p$ are large, while the interactions
are small for the local triads $k\approx p\approx q$. This result
shows that nonlinear interactions in \emph{incompressible} fluid turbulence
is \emph{nonlocal in Fourier space as well.}

The shell-to-shell energy transfer rates have been investigated by
many researchers and ourselves. It is a common wisdom that the shell-to-shell
energy transfer is local, that is, maximum energy transfer takes place
between nearest shells. We find that local shell-to-shell energy transfer
is compatible with the nonlocal triad interactions because the local
triads occupy more Fourier space volume as compared to nonlocal ($k\approx q\gg p$)
ones. The local shell-to-shell energy transfer via nonlocal triad
interactions is consistent, as seen by Domaradzki and Rogallo \cite{Doma:Local2},
Zhou \cite{Zhou:Local}, Ohkitani and Kida \cite{Ohki}, and Zhou
and Speziale \cite{Zhou:RevLocal} in their numerical simulations
and EDQNM calculations. We have ourselves computed shell-to-shell
energy transfer numerically; our theoretical results match with numerical
results very well. In this paper we show this behaviour analytically.
The role of the smallest wavenumber mode in the triad is somewhat
confusing in earlier papers. We have resolved some of these issues. 

We observe interesting behaviour in two dimensions. The shell-to-shell
energy transfer rates to the nearby shells are forward, whereas the
transfer rates to the far off shells are backward. The net effect
is a negative energy flux. This theoretical result is consistent with
Dar et al.'s numerical finding \cite{Dar:flux}. The inverse cascade
of energy is consistent with the backward nonlocal energy transfer
in mode-to-mode picture {[}$S(k|p|q)${]}. We also show that the transition
from backward energy transfer to forward transfer takes place at $d_{c}\approx2.25$. 

The contribution of local triads to the effective or renormalized
viscosity has been debated in turbulence literature. In Yakhot and
Orszag's \cite{YakhOrsz} renormalization theory, the renormalized
viscosity gets contribution from highly nonlocal wavenumber triads.
Kraichnan \cite{Krai:local,Krai:YO} first raised the above objection,
and proposed some alternatives. The local energy transfer and and
nonlocal triad interaction results discussed in this paper could be
of relevance for this issue; this aspect needs further investigation.

To conclude, an application of field-theoretic techniques to turbulence
yields interesting results regarding triad interactions and shell-to-shell
energy transfers. The method described here has also been applied
to magnetohydrodynamic turbulence, and the results are presented in
Verma et al. \cite{Ayye:MHD}. Further investigations of locality
in compressible turbulence, and other areas of turbulence will provide
us useful clues in furthering our understanding of turbulence. 

\bibliographystyle{apsrev}

\end{document}